\documentclass[12pt]{iopart}
\usepackage{graphicx}
\begin{document}

\title[Phase glass and zero-temperature phase transition...]{Phase glass and 
zero-temperature phase transition in a randomly frustrated 
two-dimensional quantum rotor model}

\author
{Lei-Han Tang}
\address{Department of Physics, Hong Kong Baptist University,
Kowloon Tong, Kowloon, Hong Kong SAR, China }
\author
{Qing-Hu Chen}
\address{Department of Physics, Zhejiang University,
Hangzhou 310027, Zhejiang, China }

\date{\today}

\begin{abstract}
The ground state of the quantum rotor model in two
dimensions with random phase frustration is investigated.
Extensive Monte Carlo simulations are performed on the corresponding
(2+1)-dimensional classical model under the entropic sampling scheme.
For weak quantum fluctuation, the system is found to be in a 
phase glass phase characterized by a finite compressibility and
a finite value for the Edwards-Anderson order parameter, signifying
long-ranged phase rigidity in both spatial and imaginary time directions.
Scaling properties of the model near the transition to the gapped,
Mott insulator state with vanishing compressibility are analyzed.
At the quantum critical point, the dynamic exponent 
$z_{\rm dyn}\simeq 1.17$ is greater than one. Correlation length
exponents in the spatial and imaginary time directions are given by
$\nu\simeq 0.73$ and $\nu_z\simeq 0.85$, respectively, both assume
values greater than 0.6723 of the pure case. We speculate that
the phase glass phase is superconducting rather than metallic
in the zero current limit.

\end{abstract}



\maketitle

\section{Introduction}

Understanding the macroscopic state of a system of 
interacting bosons at zero temperature is of 
interest to the solution of a number of problems in 
condensed matter physics\cite{Sondhi97,Sachdev99,Newrock,Leggett01}.
The theory of superconductivity is built on Cooper pairs which can 
be treated as bosons. Mapping of flux-lines in 
type-II superconductors to a two-dimensional (2D) bosonic system has 
led to a deeper understanding of the I-V 
characteristics of cuprate superconductors\cite{Nelson88,Blatter94}. 
More recently, the realization of Bose-Einstein 
condensation (BEC) in dilute atomic alkali gases has 
provided an experimental means to systematically explore
various types of macroscopic quantum states and transitions
between them, enriching our knowledge about equilibrium and
dynamic properties of strongly correlated quantum 
systems\cite{Cooper,Ho02,Morsch}.

Previously, it has been established that repulsive bosons
in restricted geometries (such as those confined by an optical
lattice) may undergo a zero-temperature quantum phase transition
from a superfluid state to a Mott insulator state as the strength
of the interaction is increased\cite{FWGF89}. 
The superfluid state is characterized by its well-known off-diagonal
long-range order and a finite phase rigidity, whereas the Mott insulator
state is characterized by zero compressibility and gapped particle
excitations. Transition between the two gives over to
intervening states (e.g., Bose glass or Mott glass)
when disorder, either in the form of random on-site potential or random 
hopping coefficients are 
introduced\cite{FWGF89,Wallin94,Prokofev04,Sengupta07}.

In this paper we focus on a different type of disorder, i.e.,
random phase frustration on the ground state properties of a 2D
interacting bosonic system. Such a situation has been realized 
experimentally in positionally disordered Josephson junction
arrays, where the frustration can be tuned by varying the strength
of a transverse magnetic field\cite{Newrock,Yun06}. At sufficiently 
strong disorder, it has been suggested that the superfluid state changes 
into a new state of matter, the ``phase glass'' phase\cite{Phillips02}. 
Although the existence of such a state is not much disputed, its precise
physical properties have not been well established\cite{Phillips02,ryu}.

The classical limit of the above problem can be described by a 
2D XY model with random phase shifts\cite{rub}.
The phase diagram of the classical model has been worked out
by Nattermann {\it et al.}\cite{nat} using renormalization group methods.
When the frustration is weak, only a small number of localized and
tightly bound vortex-antivortex pairs
are present in the classical ground state, so that long-ranged phase
order is preserved at sufficiently low temperatures. 
However, as the frustration exceeds a certain critical
value, the ground state becomes unstable against free
vortex excitations due to a large distance instability\cite{nat,kor,cha,tang}.
Consequently, vortices and antivortices
proliferate and destroy the ordered state at all temperatures.

The gauge glass model represents an extreme case where the random
frustration attains its maximum strength\cite{Ebner,FTY}. 
Despite extensive studies by a number of 
groups\cite{FTY,RY93,Akino,Katzgraber,NW04,CP99,Kim,HKM},
the low temperature properties of the 2D model are still controversial.
At the heart of the discussion is whether a certain form of glass 
rigidity survives despite the presence of a finite density of free
vortices in the system. Numerical simulations of the gauge glass
model have yielded contradictory conclusions with regard to a finite
temperature glass transition. In Ref. \cite{tang05},  
we have carried out an explicit analysis of vortex configurations in the 
ground and low-lying excited states of a corresponding Coulomb gas model,
where the random frustration is represented by a set of randomly
oriented dipoles which interact with the vortices\cite{nat,tang}.
The main conclusion of the analysis is that, unlike the case of an Ising 
spin glass, the ground state of the 2D Coulomb gas in the random dipolar
field is quite unique. The low-lying excited
vortex states can be described in terms of a dilute gas of localized
vortex-antivortex pairs super-imposed on a complex and critical
but otherwise innocent ground state vortex configuration. Due to the 
disorder, the density of states of these pairs 
is finite at zero excitation energy. Consequently, these pairs are 
able to participate in dielectric screening under thermal equilibrium 
conditions. Renormalization group arguments show that the ground state 
of the gauge glass is a critical state with a phase rigidity that decays
algebraically with distance. Thermal excitations of low energy
vortex-antivortex pairs at temperature $T$ lead to a finite correlation 
length $\xi(T)$ which diverges as $T$ tends to zero.
These predictions are confirmed by direct Monte Carlo simulations
of the gauge glass model.

Since the power-law decay of phase rigidity with distance $r$ requires
dielectric screening by excited vortex-antivortex pairs of size comparable 
to $r$ and the corresponding vortex movement at smaller scales, it is 
legitimate to ask if such a complex relaxation process can be realized in 
dynamic simulations and experiments. The slow dynamics for the creation and 
annhilation of distant vortex-antivortex pairs may indeed give rise to an 
apparent phase rigidity bigger than its equilibrium value. 
It is plausible that this type of glassy behavior is responsible for 
the previously reported finite temperature 
transition\cite{Yun06,CP99,Kim,HKM},
though one needs to work out the relevant energy and time scales
for a detailed verification of this scenario.

Quantum fluctuations of the phase lower the core energy 
of vortices and antivortices and facilitate their delocalization
through quantum tunneling. Sufficiently strong fluctuations of
this type destroy long-ranged phase ordering as in the pure case,
giving rise to the Mott insulator state. In the following we present
detailed numerical results to show that, when the fluctuations
are weak, the ground state of the system retains phase rigidity both 
in space and in time as in the classical case. This phase glass
phase is characterized by a finite Edwards-Anderson order parameter,
a finite compressibility, but a vanishing superfluid density or
helicity modulus. We also carry out a scaling analysis of the
transition to the Mott insulator state. Our results for various
critical exponents are significantly different from those of the
pure case, suggesting that the phase glass to the Mott insulator transition
belongs to a different universality class than the three-dimensional (3D)
XY model\cite{Sondhi97,Sachdev99,FWGF89}.

The paper is organized as follows. In Sec. 2, we introduce the
model Hamiltonian and the mapping to the (2+1)-dimensional classical model.
The procedure for performing Monte Carlo simulations under an entropic
sampling scheme is outlined. Various quantities characterizing
ordering in the phase glass phase are introduced. We also review
briefly a general scaling theory developed by Fisher {\it et al.}\cite{FWGF89}
for the analysis of the critical behavior at the quantum phase transition
of a bosonic system. Section 3 contains results from extensive Monte
Carlo simulations of the (2+1)-dimensional model. The critical
exponents are extracted based on a finite-size scaling analysis.
Significance of these findings are discussed briefly in Sec. 4.
The mapping between the 2D quantum model and the (2+1)-dimensional
classical model is presented in the Appendix for easy reference.

\section{The randomly frustrated quantum rotor model and its 
basic properties}

\subsection{The Hamiltonian}

We consider a 2D Josephson junction array of superconducting grains
in a transverse magnetic field. In a coarse-grained description, 
the Hamiltonian can be written as,
\begin{equation}
\hat H={E_Q\over 2}\sum_{i}\hat n_i^2-E_J\sum_{\langle ij\rangle}
\cos(\theta_i-\theta_j-a_{ij}).
\label{Hamiltonian}
\end{equation}
Here $\hat n_i$ and $\theta_i$ are the number operator of (excess)
Cooper pairs and the phase of the superconducting order parameter on
grain $i$, respectively. They satisfy the commutation relation
$[\theta_j,\hat n_k]=i\delta_{j,k}$. In the $\theta$-representation,
we may write $\hat n_j=-i\partial/\partial\theta_j$. 
The first term on the right-hand-side of (\ref{Hamiltonian}), which sums
over all $N$ sites of a square lattice, represents on-site Coulomb repulsion
between the Cooper pairs whose strength is specified by the charging
energy $E_Q=4e^2/C$, with $C$ being the capacitance of a single grain.
The second term represents Josephson coupling between neighboring islands
with strength $E_J$, and the sum is over all nearest neighbor bonds of 
the lattice.

The phase shifts $a_{ij}$ in Eq. (\ref{Hamiltonian})
are related to the vector potential ${\bf A}$ of an external magnetic field
through,
\begin{equation}
a_{i,j}=\frac{2\pi}{\Phi_0}\int_{i\rightarrow j}{\bf A}\cdot d{\bf l}.
\end{equation}
Here ${\Phi_0}=hc/(2e)$ is the elementary flux quantum.
In the present paper we shall focus on the maximally frustrated
case where the random phase shifts $a_{ij}$'s are uniformly distributed
on the interval $[0,2\pi)$ and uncorrelated from bond to bond.
The case $a_{ij}=0$ corresponds to the well-known quantum rotor 
model\cite{Sachdev99}.

The Hamiltonian (\ref{Hamiltonian}) is invariant under a global rotation,
$\theta_i\rightarrow\theta_i +c$, all $i$. This symmetry is important in
the discussion of the low energy excitations of the system.

\subsection{Mapping to a (2+1)-dimensional classical model}

It is well-known that, as far as the thermal equilibrium properties
are concerned, the quantum rotor model can be mapped to a (2+1)-dimensional
classical system using the Trotter formula\cite{Sondhi97,Wallin94}. 
This mapping forms the basis of our quantum Monte Carlo calculation.

Let us start with the partition function
\begin{equation}
Z(\beta)={\rm Tr\ }\exp(-\beta \hat H),
\label{partition-fn-q}
\end{equation}
where $\beta=1/(k_BT)$ is the inverse temperature.
Following the procedure described in Appendix A, we arrive at
a classical model defined by the action (\ref{path_H}).
The model can be brought into a dimensionless form through
the introduction of $z=\tau/\tau_0$ as the coordinate in the 
third direction, where $\tau_0=\hbar/\sqrt{E_QE_J}$.
After the transformation, we obtain
\begin{equation}
Z(\beta)=\int[{\cal D}\theta] \exp\Bigl[-{1\over K}\int_0^{L_z} dz H_P
\Bigr], \label{partition2}
\end{equation}
where $L_z=\beta\hbar/\tau_0=\sqrt{E_JE_Q}/k_BT$, 
$K=\sqrt{E_Q/E_J}$, and
\begin{equation}
H_P={1\over 2}\sum_j\Bigl({d\theta_j\over dz}\Bigr)^2
-\sum_{\langle ij\rangle}\cos(\theta_i-\theta_j-a_{ij}). 
\label{path_H1}
\end{equation}

In our Monte Carlo simulations of the quantum rotor model, we choose
a discretization along the $z$-axis with $dz=1$ and approximate
$d\theta_j/dz$ by $\theta_j(z+1)-\theta_j(z)$. This procedure appears to 
be rather adequate for the exploration of the phase glass phase and
the transition to the Mott insulator state. With this choice of $dz$, 
the critical point $K_c$ 
of the discrete model is expected to be somewhat different from that of
the original model, though we believe the large-distance properties
are not affected. The boundary conditions
in the $xy$-plane are chosen to be periodic, while Eq. (\ref{bc}) is used
along the $z$-direction. With these specifications, we obtain a classical
(2+1)-dimensional lattice model where $K$ plays the role of
a ``quantum temperature''. As such, the model can be simulated using the
entropic sampling scheme\cite{berg}. Since the algorithm allows the system 
to explore configurations over a broad range of values of the classical 
action (\ref{path_H1}) in a single simulation, it has a better chance to 
generate statistically independent samples for the calculation of thermal 
averages, which may become a problem for conventional Monte Carlo methods 
in the glass phase. In addition, once equilibrated, values of any measurable 
quantity over a broad range of $K$ values can be readily calculated.
This is particularly useful for analyzing the zero-temperature 
quantum phase transition. More details about our implementation of the 
scheme can be found in Ref. \cite{tang_mcmc}.

\subsection{Helicity modulus and compressibility}

Helicity modulus can be related to the superfluid density and hence
provides a direct measure of the superconducting order. It is defined by
considering the change of the free energy under a phase twist across
the system. Quite generally, such a twist modifies the action 
to $\tilde H_P=H_P+\delta H_P$. The change in the free energy,
to the second order in $\delta H_P$, is given by,
\begin{eqnarray}
\delta F&=&-k_BT\ln(\tilde Z/Z)\nonumber\\
&=&-k_BT\ln\Bigl\langle\exp(-{1\over K}\int_0^{L_z}dz
\delta H_P)\Bigr\rangle\nonumber\\
&\simeq& E_J\langle\delta H_P\rangle-{E_J^2\over 2k_BT}
\Bigl[\Bigl\langle\Bigl({1\over L_z}\int_0^{L_z}dz\delta H_P
\Bigr)^2\Bigr\rangle- \langle\delta H_P\rangle^2\Bigr].
\label{deltaF}
\end{eqnarray}
Here $\langle\cdot\rangle$ denotes thermal average.
For a twist $\Delta_x$ per bond in the $x$-direction, we have
\begin{eqnarray}
\delta H_P&=&\Delta_x\sum_{i}\sin(\theta_{i+x}-\theta_i + a_{i,x})
+{1\over 2}\Delta_x^2\sum_{i}\cos(\theta_i-\theta_{i+x}-a_{i,x})+\ldots 
\label{twist_H}
\end{eqnarray}
Writing $\delta F={1\over 2}N\rho_sE_J\Delta_x^2$, we obtain the
following expression for the helicity modulus,
\begin{eqnarray}
\rho_s&=&\langle\overline{\cos(\theta_x-\theta_{i+x}-a_{i,x})}\rangle
-{N\over K}\int_0^{L_z}dz\bigl[\langle \bar{J}_x(0)\bar{J}_x(z)\rangle
-\langle \bar{J}_x(0)\rangle^2\bigr].
\label{helicity0}
\end{eqnarray}
Here $\bar{J}_x(z)=N^{-1}\sum_i \sin(\theta_{i+x}(z)-\theta_i(z)+a_{i,x})$
is the average current along the $x$-direction in layer $z$,
and the overline bar denotes spatial average.

The compressibility $\kappa$ is related to the change in free energy for 
a phase twist $\Delta$ in the $z$-direction\cite{FWGF89}. 
The corresponding change in $H_P$ is given by,
\begin{equation}
\delta H_p=\Delta\sum_i {d\theta_i\over dz}+{1\over 2}N\Delta^2.
\label{deltahp}
\end{equation}
Inserting Eq. (\ref{deltahp}) into Eq. (\ref{deltaF}), and noting that
$\int_0^{L_z}dz (d\theta_i/dz)=\theta_i(L_z)-\theta_i(0)=2\pi n_i$,
where $n_i$ is the number of turns the angle on site $i$ makes along the
$z$-direction, we obtain
\begin{equation}
\delta F\simeq {1\over 2}NE_J\Delta^2-{E_J^2\over 2k_BT}
\Bigl\langle\Bigl({1\over L_z}\sum_i 2\pi n_i\Bigr)^2\Bigr\rangle
={1\over 2}NE_J\kappa\Delta^2,
\label{deltaF2}
\end{equation}
where
\begin{equation}
\kappa=1-{4\pi^2\over KN L_z}\Bigl\langle\Bigl(\sum_i n_i\Bigr)^2\Bigr\rangle.
\label{chi}
\end{equation}
Here $\langle n_i\rangle=0$ due to symmetry.

\subsection{The Edwards-Anderson order parameter}

The Edwards-Anderson order parameter for the quantum rotor model
can be defined via
\begin{equation}
q_{\rm EA}=\lim_{t\rightarrow\infty}
\overline{\langle e^{i[\theta_j(t)-\theta_j(0)]}\rangle},
\label{q_EA}
\end{equation}
where $e^{i\theta_j(t)}=e^{-i\hat H t/\hbar}e^{i\theta_j}
e^{i\hat Ht/\hbar}$. 

Consider now the auto-correlation function in imaginery time,
\begin{eqnarray}
C(\tau,\beta)&=&\overline{\langle e^{\hat H\tau/\hbar}e^{i\theta_j}
e^{-\hat H\tau/\hbar}e^{-i\theta_j}\rangle}\nonumber\\
&=&{1\over Z}{\rm Tr }\bigl[\overline{
e^{-(\beta\hbar-\tau)\hat H/\hbar}e^{i\theta_j}
e^{-\hat H\tau/\hbar}e^{-i\theta_j}}\bigr]\nonumber\\
&=&\overline{\langle e^{i[\theta_j(z)-\theta_j(0)]}\rangle},
\label{C-auto}
\end{eqnarray}
where $z=\tau/\tau_0$, and the last average is carried out in the
(2+1)-dimensional classical ensemble.
At $T=0$, the long-time limit $t\rightarrow\infty$ can be replaced by
the limit $\tau\rightarrow\infty$. Hence we may write,
\begin{equation}
q_{\rm EA}=\lim_{L_z\rightarrow\infty}C(L_z/2,L_z).
\label{q_EA1}
\end{equation}

\subsection{Scaling properties near the Mott insulator transition}

The Mott insulator state corresponds to the ``high temperature'' phase
of the (2+1)-dimensional classical model where quantum phase fluctuations
lead to a vanishing superfluid density, vanishing compressibility, and
an exponentially decaying phase-correlation function in imaginary time.
Due to the anisotropic form of (\ref{path_H1}),
two different lengths $\xi$ and $\xi_z$ are 
needed to describe the spatial and temporal correlations in the system.
As $K$ approaches its critical value $K_c$ at the transition from the
Mott insulator to the phase glass, both quantities are expected to diverge as
\begin{equation}
\xi\sim |K-K_c|^{-\nu},\qquad \xi_z\sim |K-K_c|^{-\nu_z},
\label{xi}
\end{equation}
where $\nu$ and $\nu_z$ are the respective exponents.
Since $\xi_z^{-1}$ corresponds to a characteristic energy or frequency scale
in the quantum rotor model, the ratio $z_{\rm dyn}=\nu_z/\nu$
defines the dynamical exponent at the transition. In the phase glass phase,
both $\xi$ and $\xi_z$ are expected to be infinite.
The exponent $z_{\rm dyn}^{\rm PG}$ of the phase glass phase, which may 
be different from its value at the transition, can be determined from 
suitable finite-size scaling properties.

Earlier, Fisher {\it et al}.\cite{FWGF89} proposed a general scaling theory 
for the quantum phase transition in bosonic systems. In particular,
based on the assumption that the singular part 
of the free energy (or ground state energy in the quantum model) in a 
correlated volume $\xi^d\xi_z$ is of order $K_c$, they determined the 
scaling dimensions of the superfluid density and compressibility in 
the transition region,
\begin{equation}
\rho_s\sim |K-K_c|^\zeta,\qquad \kappa\sim |K-K_c|^{\zeta_z},
\label{rho-scaling}
\end{equation}
where $d$ is the spatial dimension, and
\begin{equation}
\zeta=(d-2)\nu+\nu_z,\qquad \zeta_z=d\nu-\nu_z
\label{zetas}
\end{equation}
are the scaling exponents.

The auto-correlation function (\ref{C-auto}), on the other hand,
is expected to decay as a power law at $K=K_c$,
\begin{equation}
C(\tau,\infty)\sim\tau^{-1-(d-2+\eta)/z_{\rm dyn}},
\label{C-auto-critical}
\end{equation}
where $\eta$ is a new exponent. Scaling arguments then yield,
\begin{equation}
q_{\rm EA}\sim |K-K_c|^{(d-2+\eta)\nu+\nu_z}.
\label{q-EA-scaling}
\end{equation}

For the (2+1)-dimensional classical model, the ``specific heat''
$c_V(K)=-K\partial^2 f/\partial K^2$ with $f=F/(NL_z)$ is expected
to exhibit singular behavior at $K_c$. From the above assumption
for the singular part of the free energy $f_s\sim \xi^{-d}\xi_z^{-1}$,
one obtains the specific heat exponent,
\begin{equation}
\alpha=2-d\nu-\nu_z.
\label{specific-heat}
\end{equation}

In the pure case $a_{ij}=0$ and $d=2$, the quantum rotor model
is mapped to the 3D classical XY model.
Consequently $z_{\rm dyn}=1$, and $\zeta=\zeta_z=\nu=\nu_z$,
i.e., $\rho_s$ and $\kappa$ scale the same way as $\xi_z^{-1}$
which provides the only energy scale of the problem. Numerical 
calculations have yielded $\nu_{\rm 3DXY}\simeq 0.6723$\cite{Cucc}.
From the scaling relation $\gamma=(2-\eta)\nu\simeq 1.319$ we 
obtain $\eta_{\rm 3DXY}\simeq 0.04$.
The exponent $\alpha_{\rm 3DXY}\simeq -0.017$ is also very small.
One of our numerical tasks below is to check whether the same set of 
exponent values apply to the phase glass to the Mott insulator 
transition.

\section{Simulation results}

We have carried out extensive Monte Carlo simulations of the 
(2+1)-dimensional classical model (\ref{path_H1}) 
under the entropic sampling scheme.
The system is chosen to be a cubic lattice of $L_z$ layers
each containing $N=L^2$ sites in the $xy$-plane.
Test runs were performed on the unfrustrated quantum rotor model
(referrd to below as the pure model)
which generated results in good agreement with previous studies
on the superfluid to the Mott insulator transition at 
$K_{c0}\simeq 2.55$. Unless explicitly stated, the data presented
below for the disordered case are obtained from averages over 30 to 100
samples at any given size. This seems to be sufficient for illustrating
the behavior of the phase glass phase and for determining the critical
exponents of the transition within the limit set by the system
size we are able to investigate.

Figure 1(a) shows the ``specific heat'' data for the pure model
against $K$ for six different system sizes. The cusp singularity
at $K_{c0}$ with a very small exponent $\alpha$ is evident.
In comparison, as seen in Fig. 1(b), the singularity is much weaker 
in the randomly frustrated model. The dashed line in the figure with 
\begin{equation}
\alpha\simeq -0.3
\label{alpha}
\end{equation}
indicates a possible behavior in the infinite size limit that is 
consistent with our data, though in general the critical
amplitudes on the two sides of the transition need not be the same.

\begin{figure}
\includegraphics[width=14cm,clip=true]{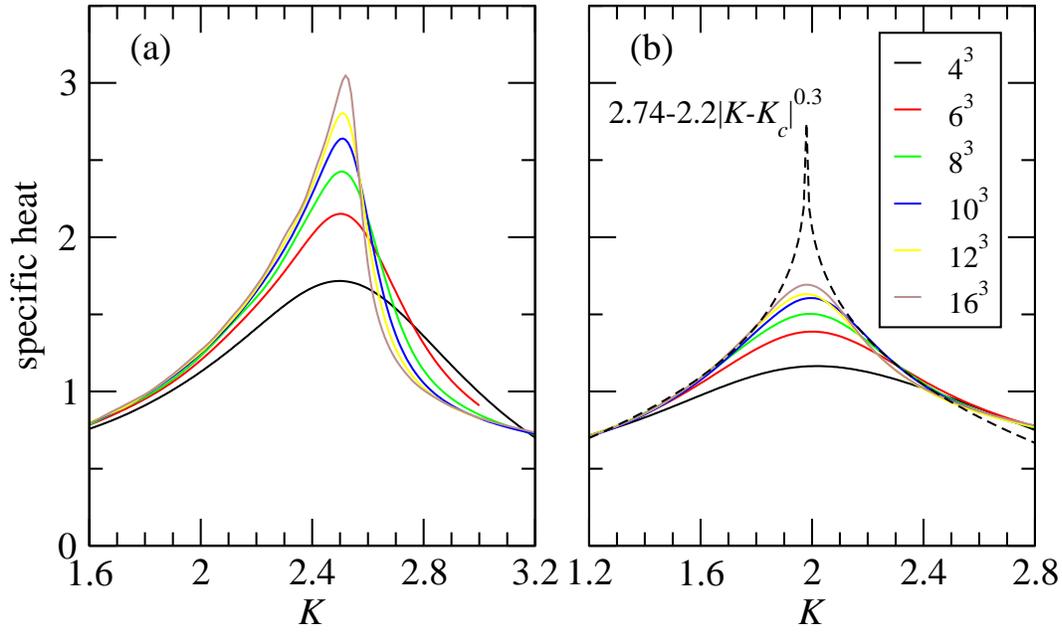}
\caption{The ``heat capacity'' against $K=\sqrt{E_Q/E_J}$ for
(a) The pure model and (b) the randomly frustrated model at
six different system sizes as indicated in the figure.
The dashed line in (b) represents a possible behavior at
infinite system size. Here $K_c=1.98$.}
\end{figure}

Figure 2 shows the helicity modulus $\rho_s$ and the compressibility
$\kappa$ of a $16^3$ system
against $K$ for four different disorder realizations.
Both quantities become vanishingly small when $K>K_c\simeq 1.98$.
At smaller values of $K$, $\rho_s$ is strongly sample-dependent
and takes on both positive and negative values. For a given sample,
$\rho_s$ may also be a non-monotonic function of $K$. The disorder-averaged 
value of $\rho_s$, on the other hand, remains close to zero.
A nonzero $\rho_s$ for individual samples signifies ``freezing'' 
of vortex loops that enables long-ranged
phase rigidity to develop. To understand the origin of negative values 
for $\rho_s$, one may consider a more general twist boundary condition 
in each layer, $\theta_{x+L,y}=\theta_{x,y}+\Delta_x$ and
$\theta_{x,y+L}=\theta_{x,y}+\Delta_y$ (see, e.g., Ref.\cite{Akino}).
Due to the random phase shifts, the minimum of the free energy
is in general achieved at some nonzero values of $\Delta_x$ and $\Delta_y$.
Depending on the particular choice of the random phase shifts,
the curvature $\rho_s$ of the free energy at $\Delta_x=\Delta_y=0$
may take on positive or negative values.
A sign change can occur when one or more vortices are relocated
on the scale of the system size. Such a move does not alter
the local phase gradients in a significant way and hence costs only a small
amount of energy. As $K$ is varied, one may envisage a change in the
relative statistical significance of configurations that differ in
this way, leading to the observed non-monotonic behavior.

In contrast, the compressibilities of the four samples differ only
slightly from one another, with no significant broadening near $K_c$.
Other quantities, such as the specific heat and correlation functions,
show the same behavior. Therefore the system is expected to be 
self-averaging in the large-size limit.

\begin{figure}
\includegraphics[width=14cm,clip=true]{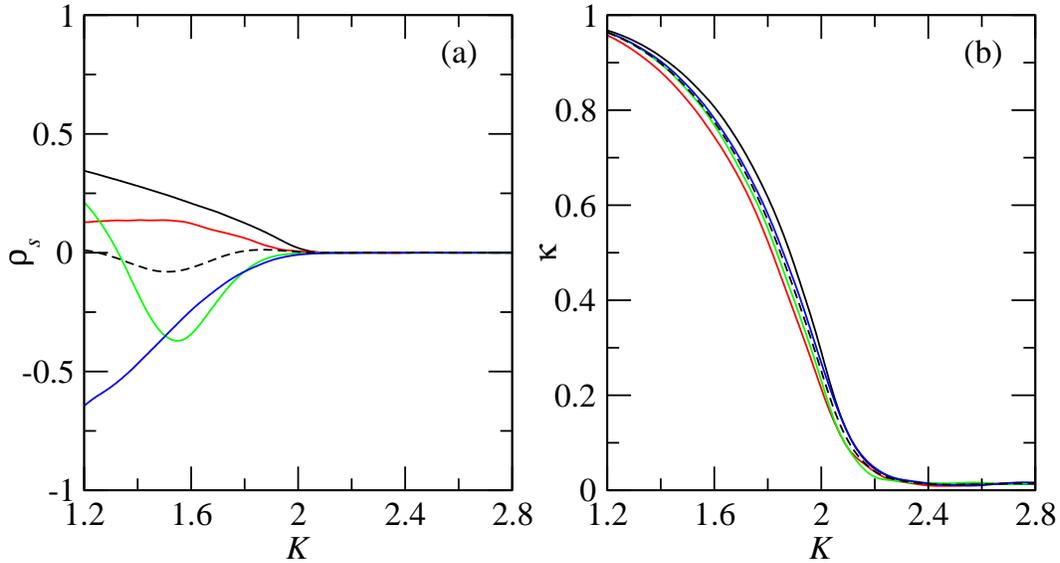}
\caption{(a) The helicity modulus and (b) the compressibility against 
$K=\sqrt{E_Q/E_J}$
for four disorder realizations at $L=L_z=16$. The dashed line in
each case indicates average over the four samples.}
\end{figure}

The critical exponents for the transition can be determined by applying
suitable finite-size scaling forms. Below we shall perform the analysis
using $L_z$ as the scaling variable. The general finite-size scaling
ansatz of a quantity $X$ then reads,
\begin{equation}
X(K,L,L_z)=L_z^{-x/\nu_z}\hat X\bigl(
(K-K_c)L_z^{1/\nu_z},L^{z_{\rm dyn}}/L_z\bigr),
\label{finite-size-scaling}
\end{equation}
where $x$ is the critical exponent for the quantity $X$.
Equation (\ref{finite-size-scaling}) is quite difficult to use
in general due to the simultaneous presence of two scaled variables.
However, as we shall see below, the exponent $z_{\rm dyn}$ is
quite close to (but larger than) one so that, for the range of system 
sizes considered, the second argument is approximately constant and
does not affect significantly the value of the function.
The validity of this assumption is justified {\it a posteriori}
by the consistency of the exponent values obtained.

Figure 3(a) shows the disorder averaged compressibility $\kappa$ 
against $K$ for six different system sizes, ranging from $L=L_z=4$
to $L=L_z=16$. At $K=K_c$ (as indicated
by the dashed line in the figure), the decay of $\kappa$ against $L_z$
can be fitted to a power law with an exponent 
$\zeta_z/\nu_z= 0.7\pm 0.1$.
From the scaling relation (\ref{zetas}) we obtain
\begin{equation}
z_{\rm dyn}= 1.17\pm 0.07.
\label{z_dyn_PG}
\end{equation}
Combining this result with our previous estimate 
$\alpha=2-2\nu-\nu_z\simeq -0.3$ yields 
\begin{equation}
\nu\simeq 0.73,\qquad
\nu_z\simeq 0.85,\qquad\zeta_z=2\nu-\nu_z\simeq 0.61.
\label{exponents_PG}
\end{equation}
The scaling plot shown in
Fig. 3(b) is generated using these exponent values.
A reasonable data collapse is seen. Note that there is
a slight increase in the slope of the curves with increasing
$L$ or $L_z$, which is consistent with the expected trend associated
with a gradual increase of the scaled variable $L^{z_{\rm dyn}}/L_z$ 
for this set of data.

\begin{figure}
\includegraphics[width=14cm,clip=true]{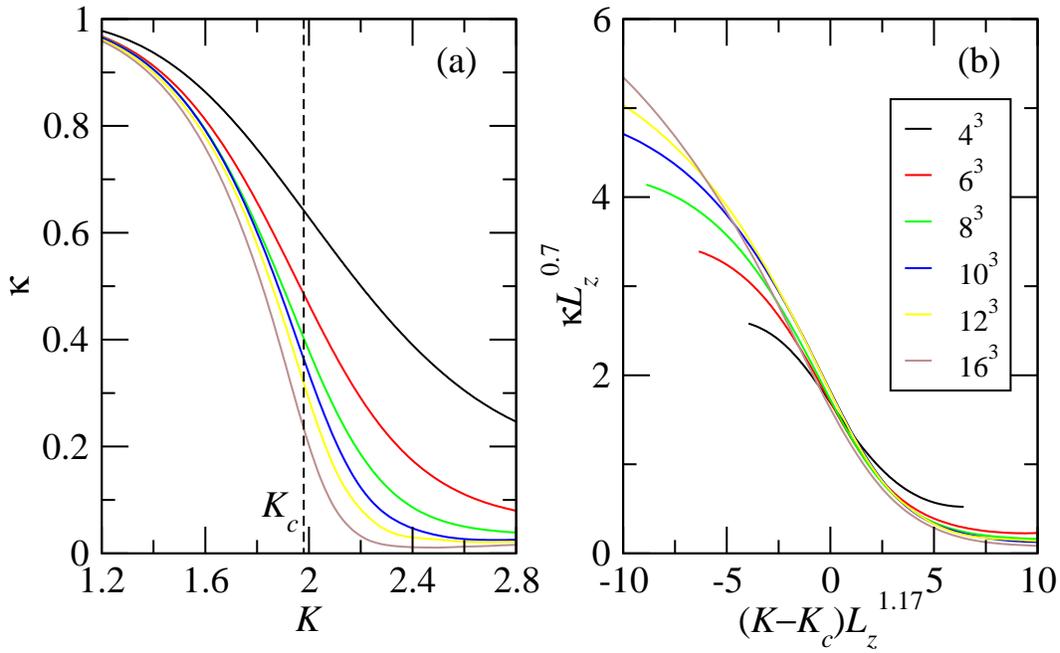}
\caption{(a) The disorder averaged compressibility for six different
system sizes $L=L_z=4,6,8,10,12,$ and 16. 
(b) A scaling plot using the exponents discussed in the text.
}
\end{figure}

We now examine the behavior of the Edwards-Anderson order parameter
$q_{\rm EA}=C(L_z/2,L_z)$. As shown in Fig. 4(a), the effect of finite 
size is more pronounced here as compared to $\kappa$ in Fig. 3(a).
Also, the dependence of $q_{\rm EA}$ on $L_z$ at a finite $L$ is
complicated by the fact that, in a finite system, the spectrum of
$\hat H$ is always discrete so that, strictly speaking, $q_{\rm EA}$
as defined by (\ref{q_EA1}) vanishes in the limit $\beta\rightarrow\infty$.
Nevertheless, for the choice $L_z=L$, we observe that data at a given
$K<K_c$ can be well fitted by a quadratic function in $1/L_z$, as
shown in Fig. 4(b). Using such an extrapolation procedure, we obtain a 
value for $q_{\rm EA}$ at each $K$ in the infinite size limit
as indicated by the dashed line in Fig. 4(a).
The approach of $q_{\rm EA}$ to zero as $K$ tends to $K_c$ can be
described by a power-law with an exponent 0.8.
Comparing with Eq. (\ref{q-EA-scaling}) and the values mentioned above
for $\nu$ and $\nu_z$, this result is consistent with a small value for
the exponent $\eta$.

\begin{figure}
\includegraphics[width=14cm,clip=true]{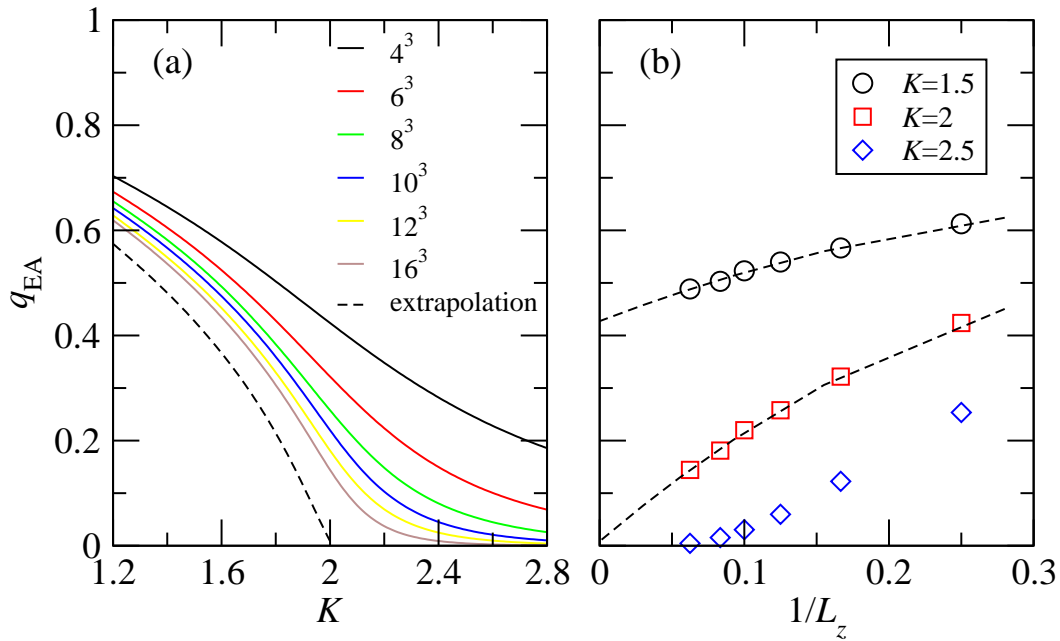}
\caption{(a) The disorder averaged Edwards-Anderson order 
parameter for six different system sizes $L=L_z=4,6,8,10,12,$ and 16. 
Dashed line indicates extrapolated value using a quadratic fit.
(b) The dependence of $q_{\rm EA}$ on $L_z$ for selected values of $K$.
}
\end{figure}

\section{Summary and discussions}

The main findings of our numerical investigation of the 
quantum rotor model at maximal random frustration are summarized
as follows. When the charging energy $E_Q$ (or boson repulsion)
is small compared to the Josephson energy $E_J$ (or boson hopping energy),
the system behaves quite similarly to the classical gauge glass model
at zero temperature. The compressibility $\kappa$ is positive and
increases with decreasing $K=\sqrt{E_Q/E_J}$. The Edwards-Anderson
order parameter $q_{\rm EA}$ also assumes a finite value, signifying 
long-ranged phase ordering in time. The $1/\tau$ (or $1/L_z$) correction 
to $q_{\rm EA}$ in imaginary time [see Fig. 4(b)] is the same as in the 
isotropic 3D XY model, suggesting a dynamic exponent $z_{\rm dyn}^{PG}=1$
and linear dispersion for the gapless phase modes.
These properties do not seem to be affected by the power-law decaying
spin-glass stiffness in the spatial direction obtained 
previously\cite{Akino,Katzgraber,tang05}.
Our results suggest that quantum tunneling between
the classically distinct low energy vortex states as discussed in
Ref.\cite{tang05} is suppressed at large distances in the phase glass.
The existence of spatial (glassy) phase order is also supported
by a nonvanishing helicity modulus $\rho_s$ below $K_c$ in individual
samples, though due to the random phase shifts, $\rho_s$ does not have
a definitive sign. These observations suggest that diffusive transport
of vortices under an applied current is unlikely in the present 
model except perhaps at $K=K_c$, casting doubt on the link 
between the phase glass and bose metal when the only quenched
disorder in the system is in the form of random phase 
frustrations\cite{Phillips02}.

We have also attempted to determine the critical exponents characterizing
the phase glass to the Mott insulator transition. Due to the relatively
small system sizes available, the estimated values of the exponents
should be considered as only tentative. With this caveat in mind,
our data are broadly consistent with a dynamic exponent 
$z_{\rm dyn}\simeq 1.17\pm 0.07$ greater than one, and correlation length
exponents $\nu\simeq 0.73$ and $\nu_z\simeq 0.85$, both greater than their
value 0.6723 in the pure case. It would be interesting to confirm
(or disprove) these results with more efficient numerical algorithms
applied to the model.

\section*{Acknowledgements} This article is dedicated to Professor Thomas
Nattermann on the occasion of his 60th Birthday. One of us (LHT) 
was affiliated with Thomas' group for six years in the early 90's.
Thomas' deep insight on ordering and phase transitions in disordered systems,
and his prowess in applying scaling arguments, with their many facets and 
subtleties, to tackle some of the most challenging problems in
statistical physics, have been a constant source of 
inspiration for people around him.
It has been a privilege to have worked closely with Thomas and
to share his vision about theoretical research.

The work is supported in part by the Research Grants Council of the 
Hong Kong SAR under grants HKBU 2017/03P and HKU3/05P, and by the 
HKBU under grant FRG/01-02/II-65. QHC acknowledges support by
the National Basic Research Program of China (Grant No. 2006CB601003).
Computations were carried out at HKBU's 
High Performance Cluster Computing Centre Supported by Dell and Intel.

\appendix
\section{Path integral representation of the partition function}

Following the standard procedure, we define
$\hat M=\exp(-\beta\hat H/n)$ and write
\begin{equation}
Z(\beta)={\rm Tr }(M^n).
\label{Z-product}
\end{equation}
When the integer $n$ is large, we may
write 
\begin{equation}
\hat M=\exp(-\beta(\hat H_Q+\hat H_J)/n)\simeq
\exp(-\beta\hat H_Q/n)\exp(-\beta\hat H_J)/n),
\label{M-matrix}
\end{equation}
where
\begin{eqnarray}
\hat H_Q&=&{E_Q\over 2}\sum_{i}\hat n_i^2=-{E_Q\over 2}\sum_{i}
{\partial^2\over\partial\theta_i^2},
\label{charging-H}\\
\hat H_J&=&-E_J\sum_{\langle ij\rangle}\cos(\theta_i-\theta_j-a_{ij}),
\label{Josephson-H}
\end{eqnarray}
are the Coulomb and Josephson energies, respectively, which do not 
commute.

Let $|\{\theta\}\rangle$ be a state where each rotor $j$ has a 
definitive phase $\theta_j\in [0,2\pi)$, and $|\{m\}\rangle$ be a state where 
each rotor $j$ has a definitive angular momentum $m_j=0,\pm 1,\pm 2,\ldots$
The matrix elements of $\hat M$ can be written as,
\begin{eqnarray}
M(\{\theta'\},\{\theta\}) &\equiv& 
\langle\{\theta'\}|\hat M|\{\theta\}\rangle\nonumber\\
&\simeq& \langle\{\theta'\}|\exp(-\beta\hat H_Q/n)\exp(-\beta\hat H_J)/n)
|\{\theta\}\rangle\nonumber\\
&=&\langle\{\theta'\}|\exp(-\beta\hat H_Q/n)
\sum_{\{m\}}|\{m\}\rangle\langle\{m\}|\exp(-\beta\hat H_J)/n)
|\{\theta\}\rangle\nonumber\\
&=&\exp\Bigl[{\beta E_J\over n}
\sum_{\langle ij\rangle}\cos(\theta_i-\theta_j-a_{ij})\Bigr]
\nonumber\\
&&\times\sum_{\{m\}}\exp\Bigl[-{\beta E_Q\over 2n}
\sum_j m_j^2+i\sum_jm_j\delta\theta_j\Bigr],
\label{M-step1}
\end{eqnarray}
where $\delta\theta_j=\theta_j'-\theta_j$ and we have used
$\langle\{\theta\}|\{m\}\rangle=\exp(i\sum_j m_j\theta_j)$.
With the help of the Poisson summation formula,
\begin{equation}
\sum_{m=0,\pm 1,\pm 2,\ldots} f(m)
=\sum_{s=0,\pm 1,\pm 2,\ldots}\int du f(u)\exp(2\pi i us), \label{Poisson}
\end{equation}
the sum over the angular momentum eigenstates $\{m\}$ can be carried out,
\begin{eqnarray}
&&\sum_{\{m\}} \exp\Bigl[-{\beta E_Q\over 2n}
\sum_j m_j^2+i\sum_jm_j\delta\theta_j\Bigr]\nonumber\\
=&&\sum_{\{s\}}\int[du]
\exp\Bigl[-{\beta E_Q\over 2n}\sum_j u_j^2
+i\sum_j u_j(\delta\theta_j+2\pi s_j)
\Bigr]\nonumber\\
=&&A\sum_{\{s\}} \exp\Bigl[-{n\over 2\beta E_Q}\sum_j
(\delta\theta_j+2\pi s_j)^2\Bigr],
\label{M-step2}
\end{eqnarray}
where $A$ is a numerical constant.

With these preparations we may carry out the matrix multiplication and
trace in Eq. (\ref{Z-product}) to obtain,
\begin{eqnarray}
Z(\beta)&=&\lim_{n\rightarrow\infty}
\int[{\cal D}\theta] \prod_{k=0}^{n-1}
M\bigl(\{\theta(\tau_{k+1})\},\{\theta(\tau_k)\} \bigr),\nonumber\\
&=&\int[{\cal D}\theta]
\exp\Bigl[-{1\over\hbar}\int_0^{\beta\hbar}d\tau H_{c}\Bigr],
\label{partition}
\end{eqnarray}
where $\tau_k=k\hbar\beta/n$ is the imaginery time coordinate and
\begin{equation}
H_c={\hbar^2\over 2E_Q}\sum_j\dot\theta_j^2
-E_J\sum_{\langle ij\rangle}\cos(\theta_i-\theta_j-a_{ij})
\label{path_H}
\end{equation}
is the classical action. The integration over the $\theta_j(\tau_k)$'s are
on the infinite domain $(-\infty,\infty)$ for all $k$ except
at $k=0$, where the domain $[0,2\pi)$ is taken instead.
This procedure takes care of the sum over the $s_j$'s as in Eq. (\ref{M-step2}).
Note that 
\begin{equation}
\theta_j(\beta\hbar)=\theta_j(0)\quad{\rm mod}\ 2\pi. 
\label{bc}
\end{equation}

\end{document}